# Reliable scaling exponent estimation of long-range correlated noise in the presence of random spikes

Radhakrishnan Nagarajan

Center on Aging, University of Arkansas for Medical Sciences

**Abstract**:

Detrended fluctuation analysis (DFA) has been used widely to determine possible long-range correlations in data obtained from diverse settings. In a recent study [1], uncorrelated random spikes superimposed on the long-range correlated noise (LR noise) were found to affect DFA scaling exponent estimates. In this brief communication, singular-value decomposition (SVD) filter is proposed to minimize the effect random spikes superimposed on LR noise, thus facilitating reliable estimation of the scaling exponents. The effectiveness of the proposed approach is demonstrated on random spikes sampled from normal and uniform distributions.

*Author for Correspondence*

Radhakrishnan Nagarajan
Center on Aging, University of Arkansas for Medical Sciences
629 Jack Stephens Drive, Room: 3105
Little Rock, Arkansas 72205
Phone: (501) 526 7461
Email: nagarajanradhakrish@uams.edu

# 1. Introduction

Detrended fluctuation analysis (DFA) is a powerful technique to determine the nature of correlations in a given data. DFA and its extensions (MF-DFA) have been used to estimate scaling exponents in a wide-range of synthetic and experimental and real-world data sets [2-4]. Experimental and real-world data sets are often corrupted with artifacts at the *dynamical* and *measurement* level. While the former is of feedback nature and coupled to the systems dynamics, the latter is an additive term superimposed on the dynamical process. Recent reports [1, 5], have indicated the susceptibility of DFA to artifacts in the form of trends and random spikes superimposed on the long-range correlated noise (LR noise). Such artifacts have been found to introduce spurious crossovers preventing reliable estimation of the scaling exponents. In [6], singular-value decomposition (SVD) filter was proposed to minimize the effect of linear, power-law, periodic and quasi-periodic trends superimposed on LR noise. It is important to note that the above trends fall under the class of *deterministic* artifacts. *Non-deterministic* artifacts such as random spikes (RS) [1] superimposed on LR noise have also been found to affect DFA scaling exponent estimates. In [1], was concluded that the fluctuation function of LR noise superimposed with RS obey the *superposition rule*. In other words, superimposing LR noise with RS significantly alters its correlation properties preventing reliable estimation of the scaling exponent. Inspired by these reports [1, 5], the SVD filter proposed in [6] is extended to minimize the effect of RS superimposed on LR noise.

The report is organized as follows, in Section 2, properties of random spikes is investigated. An algorithm based on SVD is proposed to minimize the effect of RS superimposed on LR noise. In Section 3, the effect of the SVD filter on reliable estimation of scaling exponent LR using DFA is demonstrated. RS sampled from uniform and normally distributed uncorrelated noise are considered.

**2. Minimize the effect of random spikes**

Linear transformation techniques such as SVD have been successfully used in the past to discern noise from signal in the given data [7]. In the case of deterministic trends superimposed on LR noise [1], a marked decrease in the magnitude of the eigen-values is observed separating the trend from the LR noise. In other words, deterministic trends can be approximated to a narrow band of frequencies in the broad-band LR noise [6]. This has to be contrasted with random spikes sampled from uncorrelated noise, whose energy is spread across the entire spectrum. Ideally, eigen-decomposition of uncorrelated noise embedded in an m-dimensional space yields a uniform distribution of eigen-values. From a geometrical perspective, uncorrelated noise fills the entire m-dimensional space and whose volume is maximum, representing a sphere [8]. Increasing correlations in any direction is accompanied by an increase in the magnitude of the corresponding eigen-value resulting in m-dimensional ellipsoids, Figure 1. Alternately, skewed distribution of eigen-values is indicative of possible correlations in the given data. In the present study, Morgera's covariance complexity (*h*) [9] is used to capture the skewness in the eigen-spectrum, hence the randomness of the given data. A brief description is enclosed below for completeness.

*Morgera's Covariance Complexity (h):*

**Given**: Given data $e, \{e_n\}, i = 1...N$

**Step 1**: Embed $\{e_n\}$ with parameters (*m*, *t*) [6] where *m* is the embedding dimension and *t* the time delay. The embedded data can be represented as a matrix $\Gamma$ with elements:

$$g_k = (e_k, e_{k+t}, ..., e_{k+N-(m-1)t}), 1 \leq k \leq m$$

$$\Gamma = \begin{bmatrix} g_1 \\ . \\ . \\ g_m \end{bmatrix} \qquad (1)$$

The time delay is fixed at ($\tau = 1$), therefore $e_{ij} = e_{i+j-1}$, $1 \leq i \leq N - (m-1)\tau$ and $1 \leq j \leq m$.

**Step 2**: Since the matrix is an embedding of random spike it is full rank for any choice of m. SVD of $\Gamma$, yields eigen values $l_i$, $i = 1...m$. The normalized variance is given by

$$s_i = \frac{(l_i)^2}{\sum_{i=1}^{m}(l_i)^2}, i = 1...m$$

**Step 3**: Morgera's covariance complexity ($h$) of the random spike is

$$h = -\frac{1}{\log m} \sum_{k=1}^{m} s_k \log s_k$$

The value of $h$ lies in the closed interval $(0 \leq h \leq 1)$ and is inversely proportional to the correlation in the given data.

Consider the linearly correlated noise given by $\frac{dx}{dt} = b\, x(t) + g(t)$ where $b = -0.8$ and $g(t)$ is normally distributed uncorrelated noise with zero mean and unit variance. The discrete counter part is given by $x(t + \Delta t) = (1 + b\,\Delta t)x(t) + g\,\Delta t$. Two-dimensional representation (m = 2, $\tau$ = 1) [6] of the one-dimensional time series $x$ sampled ($\Delta t = 0.02$) from the correlated noise Figure 1a, exhibits an elliptical shape reflecting skewed distribution of normalized eigen-value and ($h \sim 0$). This has to be contrasted with that of its random shuffle which exhibits uniform distribution of the eigen-values, and ($h \sim 1$), Figure 1b. Random shuffle was generated by bootstrapping $x$ without replacement [10]. While the amplitude distribution of $x$ is retained in the shuffled surrogates the correlation between the samples is destroyed. Thus random shuffles represent the uncorrelated counterpart of the given correlated data. The power-spectrum of the correlated noise its shuffled surrogate is shown in Figures 1c and d, respectively. Uniform distribution of eigen-values in the case of the shuffled surrogates is accompanied by a flat power-spectrum, whereas skewed distribution is accompanied by a decaying power-spectrum. Thus Morgera's complexity of can discern the extent of correlation in the given data.

Spikes can be generated from nonlinear deterministic as well as non-deterministic processes, and contain valuable information regarding the systems dynamics [10-12]. In the present study, we shall consider spikes to random artifacts generated according to ([13], personal communication, P. Ch. Ivanov). Random spikes (RS) sampled with ($\mu = 0$, $\sigma = 5$) from normal and uniformly distributed uncorrelated noise were superimposed on LR noise ($\sigma = 1$, $\mu = 0$, $\alpha = 0.8$). The standard deviation of the RS were forced to be significantly higher compared to that of the LR noise so as to significantly alter the correlation properties of the LR noise [13]. The spikes were generated with a specified *acceptance probability* (p) [13]. The acceptance probability can be thought of as the *mean firing rate* of a neuron [11]. We implicitly assume the acceptance probabilities of the spikes to be constant across the entire length of the data, thus *homogenous*. This has to be contrasted with cases where the acceptance probability can vary as a function of time, *heterogeneous*. It is important to note that the density of the spikes is governed by p. In [1], it was pointed out that scaling of superimposed data ($y = x + e$) consisting of the LR noise ($x$) and RS ($e$) obeys the superposition rule. The impact of spikes on the LR noise is governed mainly by two parameters, namely: its standard deviation of the ($\sigma$) and the acceptance probability (p). Thus prior to discussing the SVD filtering it is important to understand the impact of these parameters on the scaling of $y$. Random spikes were sampled from normal and uniform distributions with acceptance probability (p = 0.01, 0.05, 0.10, 0.20, 0.30, 0.40, 0.50, 0.60, 0.70, 0.80, 0.90 and 1.0) for a fixed standard deviation ($\sigma = 5$ with $\mu = 0$). These were subsequently superimposed on the LR noise ($\mu = 0$, $\sigma = 1$). The log-log of the fluctuation function versus the time scale of $y$ revealed that for low values of p, the scaling of $y$ resembled of the LR noise ($\alpha = 0.8$) whereas for high values of p, it resembled that of uncorrelated noise ($\alpha = 0.5$). This behavior was immune to the distribution of the uncorrelated noise generating the random spikes, Figures 2a and 2b. We also investigated the impact of the standard deviation ($\sigma$) of the $e$ on the scaling of $y$ for a fixed acceptance probability (p). Random spikes were sampled from normal and uniform distributions with standard deviations ($\sigma$ = 1, 5, 10, and 20 with $\mu$ = 0) for a fixed acceptance probability (p = 0.05). These were subsequently superimposed on the LR noise ($\mu$ = 0, $\sigma$ = 1). The scaling behavior of $y$ for low values of $\sigma$ resembled that of the LR noise ($\alpha$ = 0.8). However, with increasing $\sigma$ the scaling of $y$ resembled that of uncorrelated noise ($\alpha$ = 0.5). This behavior was immune to the distribution of the uncorrelated noise generating the RS, Figures 3a and 3b. Since the exponent of $x$ is fixed ($\alpha$ = 0.8) and $y = x + e$, the fluctuation function of $e$ is a function of the acceptance probability (p) and the standard deviation ($\sigma$).

In the present, study, we choose the standard deviation ($\sigma$ = 5) and acceptance probability (p = 0.05 and p = 0.10). Qualitative inspection of the waveforms indicates that the spikes with these parameters introduce marked distortion in the data, figure 4. The power spectrum of the LR noise ($x$, $\alpha$ = 0.8, N = 7168) and that superimposed with RS ($e$) sampled from normal and uniform random number generators with acceptance probabilities (p = 0.05 and p = 0.10) is shown in Figure 5. It can be observed that the power-spectrum of $y$ flattens at higher frequencies and exhibits a significant deviation from that of $x$. The extent of flattening is directly proportional to the acceptance probability. Thus the choice of the parameters ($\sigma$ = 5, p = 0.05 and p = 0.10) is valid in the present context.

SVD filtering of RS requires embedding $y$ in a high-dimensional space with embedding dimension (m) and time delay ($\tau$) [6]. As noted earlier, the density of the spikes is directly proportional to the acceptance probability. This in turn determines the sparseness of the embedding matrix. As a preliminary check, we chose to investigate the distribution of the eigen-values for the RS, LR noise and LR noise superimposed with RS with varying embedding dimensions (m = 100, 200, 300, 400, and 500) and acceptance probabilities (p = 0.05 and p = 0.10) using ($h$). The value of ($h$) estimated for the random spike ($h_e$), LR noise ($h_x$) and LR noise superimposed with random spike ($h_y$) is shown in Figure 6. It can be observed that ($h_e > h_y \gg h_x$). This inequality is immune to varying embedding dimension (m = 100, 200, 300, 400 and 500), acceptance probability (p = 0.05, p = 0.10) and choice of distributions. Thus embedding the random spikes in a high dimensional space does not affect its random nature. In order to determine the contribution of the random spike to the spectral content, the power-spectrum of LR noise $x$ ($\alpha = 0.8$, N = 7168, $\mu = 0$, $\sigma = 1$), RS ($e$) ($\mu = 0$, $\sigma = 5$, N = 7168, p = 0.05) sampled from normal distribution, and the superimposed data $y = x + e$ was investigated, Figure 7. *The most crucial observation from Figure 7, is that the contribution of the LR noise as reflected by the power-spectral magnitude decreases with increasing frequency ($1/f^a$) with negligible contribution at the higher frequencies, whereas that of the random spike persists across the entire band.* Alternately, the power in $y$ at higher frequencies is dominated by that of the random spikes. More importantly those in the frequency range ($f > f^*$), is dominated by those of random spikes. While it is not clear what would be a valid choice of $f^*$, filtering the high frequency components can minimize the effect of the random spike on the LR noise. This has to be contrasted to that of deterministic trends [6], which manifest themselves as low frequency components. From Figure 7, it can also be seen that the power spectrum of the random shuffled surrogate of $y$ resembles those of the random spikes $e$.

Ideally uncorrelated noise should exhibit uniform distribution of the eigen-values. In which case, a suitable choice would be to assign the least dominant eigen-value of $y$, as representative of $e$. However, we observed that the magnitude eigen-values of $e$ decreased with increasing embedding dimension (m), Figures 8-11. In the case of experimental data, one does not have knowledge about the probability distribution of the process generating the random spikes. Given these intricacies, we chose *scaled shuffled random surrogates* to be representative of **e**. As noted earlier, the shuffled surrogates retain the distribution of **y**. Thus no assumption is made on the nature of the process generating **e**. Through our simulation studies we observed that scaling the eigen-spectrum of the shuffled surrogates of **y** $l_i^s, i = 1...m$ with the least dominant eigen-value of **y** $l_m$ i.e. $(\frac{l_m}{l_m^s})l_i^s, i = 1...m$ resulted in a better representation of the random spike **e**, Figures 8-11. In Figure 8, the eigen-spectrum of the **e** sampled from normally distributed uncorrelated noise (μ = 0, σ = 5, p = 0.10), its random shuffled counterpart and scaled random shuffle with varying embedding dimension (m = 10, 50, 100 and 500) is shown. The eigen-spectrum of the random spike resembles that of the scaled random shuffled surrogate and deviates significantly from the least dominant eigen-value of **y** ($l_m$). It can also be observed that this deviation increases with increasing embedding dimension. This reiterates our claim that eigen-spectrum of the random spikes is not constant for a given m. A similar behavior was observed for random spikes sampled from uniformly distributed uncorrelated noise (μ = 0, σ = 5, p = 0.10), Figure 9. Increasing the acceptance probability (p = 0.90), Figures 10 and 11 minimized the discrepancy between the scaled shuffled surrogates, shuffled surrogates and the random spikes. This can be attributed to the overwhelming effect of **e** on **y** with increasing p.

**Algorithm I**

**Given**: Long-range correlated noise superimposed with random spikes generated from uncorrelated noise with unknown distribution.

**Objective**: Minimize the effect of random spikes and facilitate reliable estimation of the scaling exponent.

*Assumptions:*

  (a) *Random spikes $e$ superimposed on the long-range correlated noise $x$ is of the form $y = x + e, \{y_n\}, i = 1...N$.*

  (b) *Power-spectrum of the long-range correlated noise decays as $(1/f^a)$.*

  (c) *Random spikes $e$ and long-range correlated noise $x$ are uncorrelated.*

  (d) *Variance and acceptance probability of the random spikes is chosen so as to significantly affect the correlation properties of the long-range correlated noise at the higher frequencies or lower time scales.*

**Step 1** Embed $y$ with parameters $(m, t)$ where $m$ is the embedding dimension and $t$ the time delay [6]. The embedded data can be represented as a matrix **G** whose $k^{th}$ row is given by

$$g_k = (y_k, y_{k+t}, ..., y_{k+N-(m-1)t}), 1 \le k \le m$$

The time delay is fixed at $(t = 1)$, therefore $g_{ij} = y_{i+j-1}, 1 \le i \le N - (m-1)t$ and $1 \le j \le m$.

**Step 2** Apply SVD to the matrix **G**, to obtain $\mathbf{G} = \mathbf{U\Sigma V^T}$. Let the non-zero eigen-values in $\Sigma$ be $l_i, i = 1...m$, where $l_1 > l_1 > ... > l_m$.

**Step 3**: Generate random shuffled surrogate of $y$. Embed the shuffled surrogates with the same parameters $(m, \tau)$ as in Step 1 into $\mathbf{G^s}$. Let the non-zero eigen-values obtained by eigen-decomposition of $\mathbf{G^s}$ be $l_i^s, i = 1...m$. The corresponding scaled shuffled eigen-spectrum is

$l_i^* = (\frac{l_m}{l_m^s})l_i^s, i = 1...m$ with correlation matrix $\mathbf{R^s}$ such that $R_{ij}^s = 0$ for $i \ne j$ and $R_{ii}^s = (l_i^*)^2$.

**Step 4** The correlation matrix $\mathbf{R^F}$ of the filtered data is obtained as $\mathbf{R^F = GG^T - R^s}$. Determine the eigen-values of $\mathbf{R^F}$ $(l_i^F)^2, i = 1...m$. The corresponding embedding matrix $(\mathbf{G^F})$ of the filtered data is given by $\mathbf{G^F = U\Sigma^F V^T}$ where $\Sigma^F_{ij} = 0$ for $i \ne j$, $\Sigma^F_{ii} = l_i^F$, U and V substituted from Step 2.

**Step 5** Let the elements of $\Gamma^F$ be of the form in Step 1, such that

$$g_k^F = (y_k^F, y_{k+t}^F, ..., y_{k+N-(m-1)t}^F), \ 1 \leq k \leq m$$

The corresponding one-dimensional filtered data ($y^F$) is given by

$$y_{i+j-1}^F = g_{ij}^F \text{ where } 1 \leq i \leq N - (m-1)t \text{ and } 1 \leq j \leq m$$

## 3. Results

Random spikes $e$ sampled from uniformly and normally distributed uncorrelated noise ($\mu = 0$, $\sigma = 5$, p = 0.05 and p = 0.10) were superimposed on LR noise $x$ ($\mu = 0$, $\sigma = 1$, $\alpha = 0.8$, N = 7168) to yield $y = x + e$, Figure 12. The SVD filter Algorithm I was used to minimize the effect of spikes with (m = 500, $\tau = 1$). The power-spectrum, Figure 12, of the filtered data shows a considerable overlap with that of the original data for all the cases. This has to be contrasted with Figure 5, where the power spectrum of $x$ showed a considerable deviation from that of $y$. Thus the proposed filtering technique can minimize the effect of $e$ superimposed on $x$. The eigen-spectrum of $x$ was compared to $y$ with varying embedding dimensions (m = 100, 200, 400 and 500), Figures 14-16. Embedding dimension (m = 300) is not shown. For (p = 0.05) improved performance was observed with increasing embedding dimension Figures 13, 14. This is reflected by a considerable overlap between the eigen-spectrum of $x$ and $y$. However, for (p = 0.10) we observed considerable deviation between the eigen-spectrum of $x$ and $y$ irrespective of the choice of the embedding dimension. The proposed algorithm implicitly assumes that the random spikes are restricted to the higher frequencies in the power-spectrum. For certain choice of the standard deviation and acceptance probability the impact of random spikes is overwhelming resulting in scaling behavior similar to uncorrelated noise, Figures 2 and 3. For these parameters the power of the random spikes is distributed across the entire spectrum and not restricted to the higher frequencies. Thus the proposed algorithm fails to minimize the effect of random spikes.

The log-log plot of the fluctuation function versus the time scale of the filtered data obtained for embedding dimensions (m = 100, 200, 400 and 500) using first order DFA (DFA-1) is shown in Figures 17-20. As with earlier reports [1], random spikes superimposed on the LR noise *x* ($\mu = 0$, $\sigma = 1$, $\alpha = 0.8$, N = 7168) significantly alter the slope of the log-log plot of the fluctuation function versus the time scale, shown by the dashed lines in Figures 17-20. The slope of *y* for (p = 0.05), Figures 17 and 18, is considerably higher than for (p = 0.10), Figures 19 and 20, conforming to our earlier observation on the impact of the acceptance probability, Figures 2 and 3. The proposed Algorithm I, significantly reduces the effect of random spikes *e* ($\mu = 0$, $\sigma = 5$, p = 0.05) sampled from uniformly and normally distributed uncorrelated noise, Figures 17, 18. However, for *e* ($\mu = 0$, $\sigma = 5$, p = 0.10) significant distortion is introduced at the lower time scales indicating for these parameters the spectral content of the LR noise is dominated overwhelmingly by that of the random spikes and it might not be possible to discern the LR noise from the random spikes.

## 4. Discussion

In the present study, SVD based filter (Algorithm I) is proposed to minimize the effect of random spikes superimposed on long-range correlated noise. Random spikes were sampled from uniformly and normally distributed uncorrelated noise. The random spikes were generated according to [1, 13], and the parameters were chosen so as to significantly affect the correlation properties of the superimposed data. The proposed techniques relies on the fact that unlike random spikes which exhibits a seemingly constant spectral power, that of long-range correlated noise exhibits a ($1/f^{a}$) decay with minimal contribution at the higher frequencies. Thus it might not be possible to filter the effect of the random spikes if its contribution to the spectral content of the random spikes is overwhelming across the entire band. The proposed algorithm does not assume any particular distribution of the random process generating the spikes and uses scaled shuffled random surrogates to capture the spectral signature of the random spikes superimposed on the LR noise. The shuffled surrogates were also useful in capturing the decaying trend in the eigen-spectrum of the random spikes. Several factors influence the performance of the proposed algorithm such as the strength of the random spikes, reflected by its standard deviation and the acceptance probability. The proposed algorithm implicitly assumes that the contribution of the random spikes at the higher frequencies and hence results in poor performance in cases where the effect of random spikes is overwhelming.

## 5. Acknowledgement

I would like to thank the authors in Reference 1 for making available the data through Physionet. The present study is supported by funds from National Library of Medicine (1R03LM008853-1) and junior faculty grant from American Federation for Aging Research (AFAR).

13. Castro and T. Sauer, Reconstruction of chaotic dynamics through spike filters, [1999] Phys. Rev. E 59 (3), 2911.

13. *Personal communication P. Ch. Ivanov*

The parameters are chosen similar to Reference 1. Steps in random spike generation:

- Given: long-range correlated noise $x$ ($x_i$, $i = 1…N$) with scaling exponent $\alpha$. Make ($x_i$, $i = 1…N$) zero mean and unit variance ($\mu = 0$, $\sigma = 1$). In the present study, we chose ($\alpha = 0.8$, *available at Physionet*) and ($N = 7168$).
- Generate uncorrelated noise ($u_i$, $i = 1…N$) from a random number generator with a specified distribution. In the present study, two distributions namely: normal and uniform distributions with zero mean and unit variance are considered. Multiply ($u_i$) by a factor of five i.e. $\mu = 0$, $\sigma = 5$. It is important to note that ($e_i$) has much larger standard deviation compared to ($x_i$) For example see Figures 4 and 5.
- Generate uniform random number generator ($s_i \in [0, 1]$, $i = 1…N$). Choose the acceptance probability ($p$). In the present study, we choose $p = 0.05$ (Reference 1) and $p = 0.10$, which corresponds to 5% and 10% acceptance of the spikes $e$. i.e.

$$e_i = s_i \quad \text{if } (s_i < p)$$
$$= 0 \quad \text{otherwise}$$

- Long-range correlated noise superimposed with random spike $y$ ($y_i$, $i = 1…N$) is generated as $y_i = x_i + e_i$

**Figures and Captions**

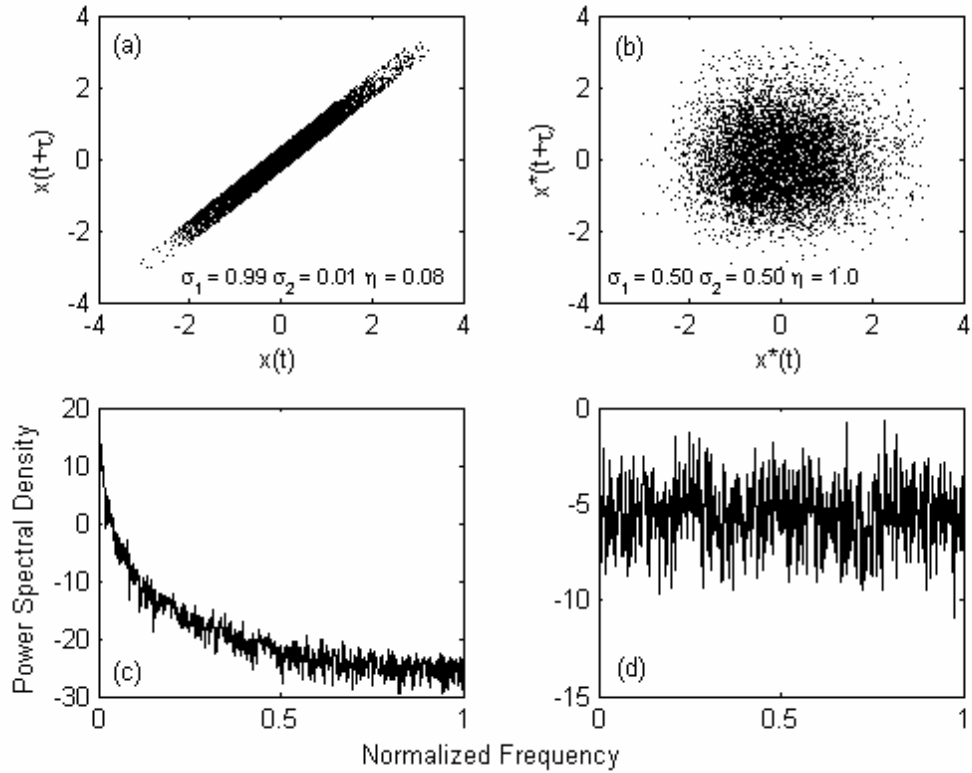

**Figure 1** Two-dimensional embedding (m = 2, τ = 1) of the linearly correlated noise and its random shuffled counterpart is shown in (a) and (b) respectively. The corresponding power-spectrum is shown in (c) and (d). The normalized variance and Morgera's covariance complexity is included in (a) and (b).

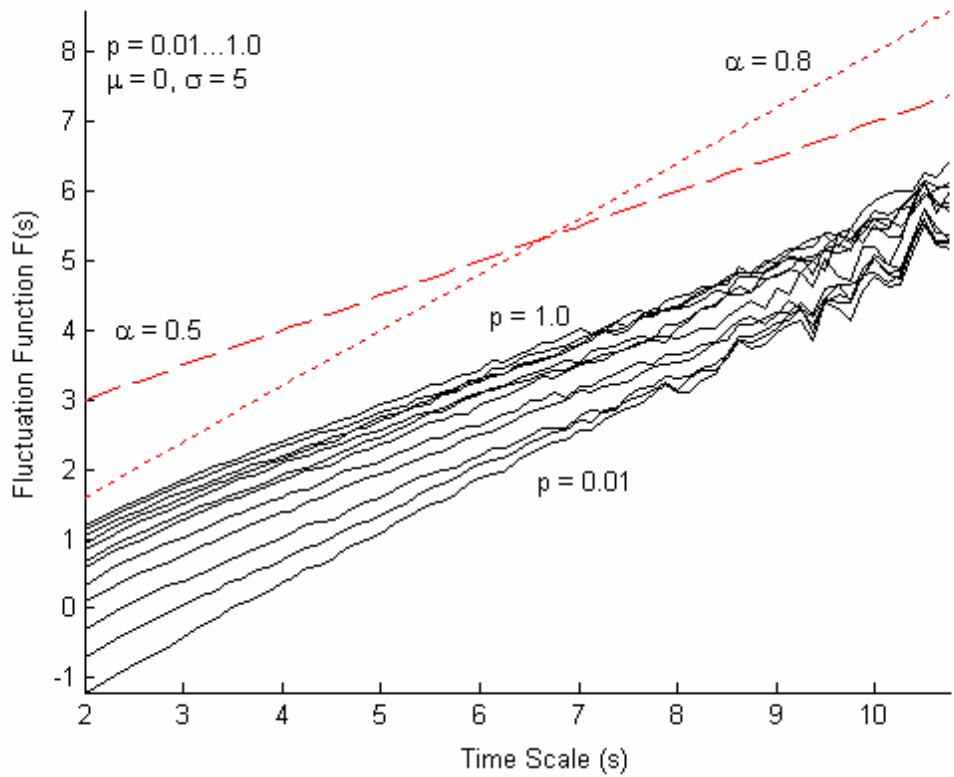

**Figure 2a** Log-log plot of the fluctuation function versus time scale of long-range correlated noise ($\alpha = 0.8$, $\mu = 0$, $\sigma = 1$, $N = 7168$) superimposed with random spikes (p = 0.01, 0.05, 0.10, 0.20, 0.30, 0.40, 0.50, 0.60, 0.70, 0.80, 0.90 and 1.0) sampled from normally distributed uncorrelated noise ($\mu = 0$, $\sigma = 5$). The acceptance probability was gradually increased from p = 0.01 (bottom most solid line) to p = 1.0 (top most solid line) in that order. The dashed and the dotted lines correspond to scaling of uncorrelated noise ($\alpha = 0.5$) and long-range correlated noise ($\alpha = 0.8$).

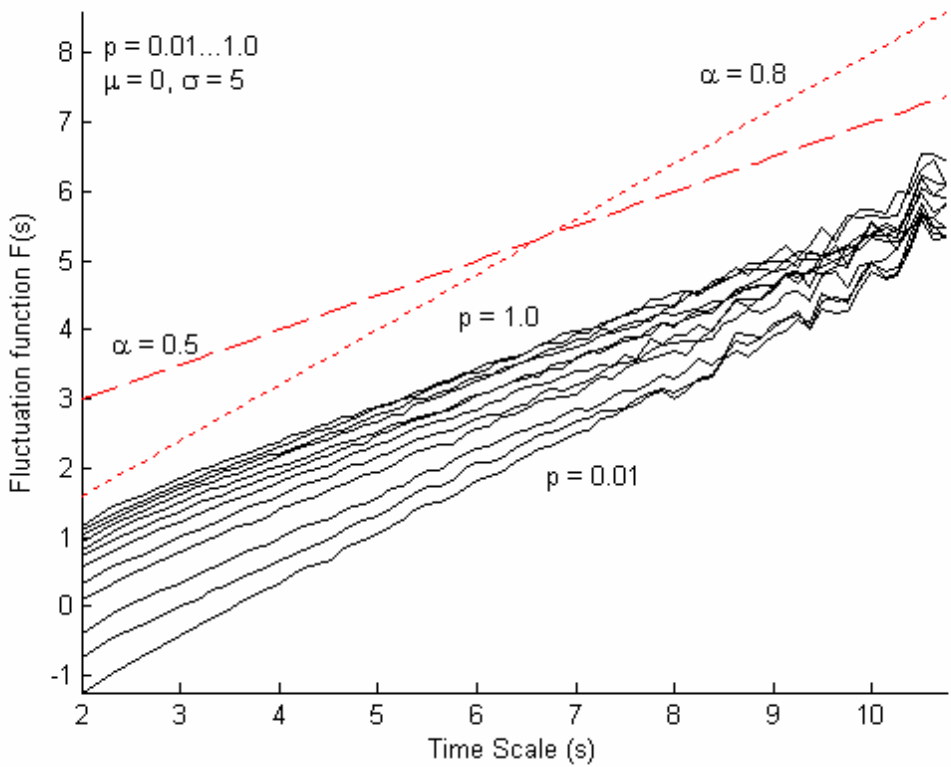

**Figure 2b** Log-log plot of the fluctuation function versus time scale of long-range correlated noise ($\alpha = 0.8$, $\mu = 0$, $\sigma = 1$, $N = 7168$) superimposed with random spikes (p = 0.01, 0.05, 0.10, 0.20, 0.30, 0.40, 0.50, 0.60, 0.70, 0.80, 0.90 and 1.0) sampled from uniformly distributed uncorrelated noise ($\mu = 0$, $\sigma = 5$). The acceptance probability was gradually increased from p = 0.01 (bottom most solid line) to p = 1.0 (top most solid line) in that order. The dashed and the dotted lines correspond to scaling of uncorrelated noise ($\alpha = 0.5$) and long-range correlated noise ($\alpha = 0.8$).

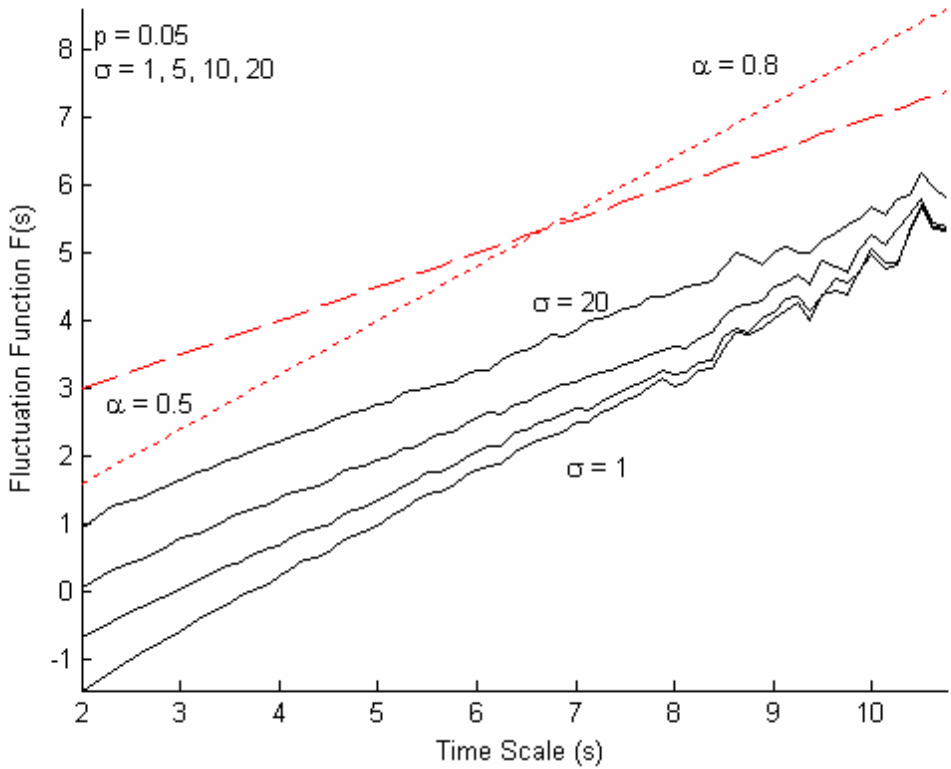

**Figure 3a** Log-log plot of the fluctuation function versus time scale of long-range correlated noise ($\alpha = 0.8$, $\mu = 0$, $\sigma = 1$, $N = 7168$) superimposed with random spikes ($\sigma = 1, 5, 10$ and $20$) sampled from normally distributed uncorrelated noise ($\mu = 0$, $p = 0.05$). The standard deviation was gradually increased from $\sigma = 1$ (bottom most solid line) to $\sigma = 20$ (top most solid line) in that order. The dashed and the dotted lines correspond to scaling of uncorrelated noise ($\alpha = 0.5$) and long-range correlated noise ($\alpha = 0.8$).

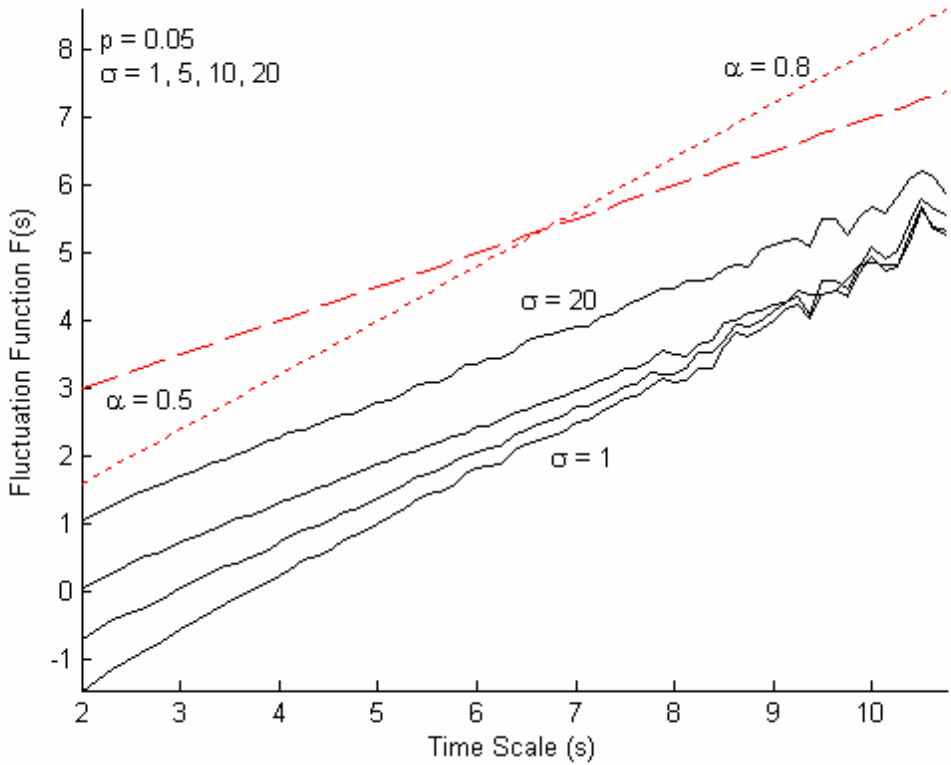

**Figure 3b** Log-log plot of the fluctuation function versus time scale of long-range correlated noise ($\alpha = 0.8$, $\mu = 0$, $\sigma = 1$, $N = 7168$) superimposed with random spikes ($\sigma = 1, 5, 10$ and $20$) sampled from uniformly distributed uncorrelated noise ($\mu = 0$, $p = 0.05$). The standard deviation was gradually increased from $\sigma = 1$ (bottom most solid line) to $\sigma = 20$ (top most solid line) in that order. The dashed and the dotted lines correspond to scaling of uncorrelated noise ($\alpha = 0.5$) and long-range correlated noise ($\alpha = 0.8$).

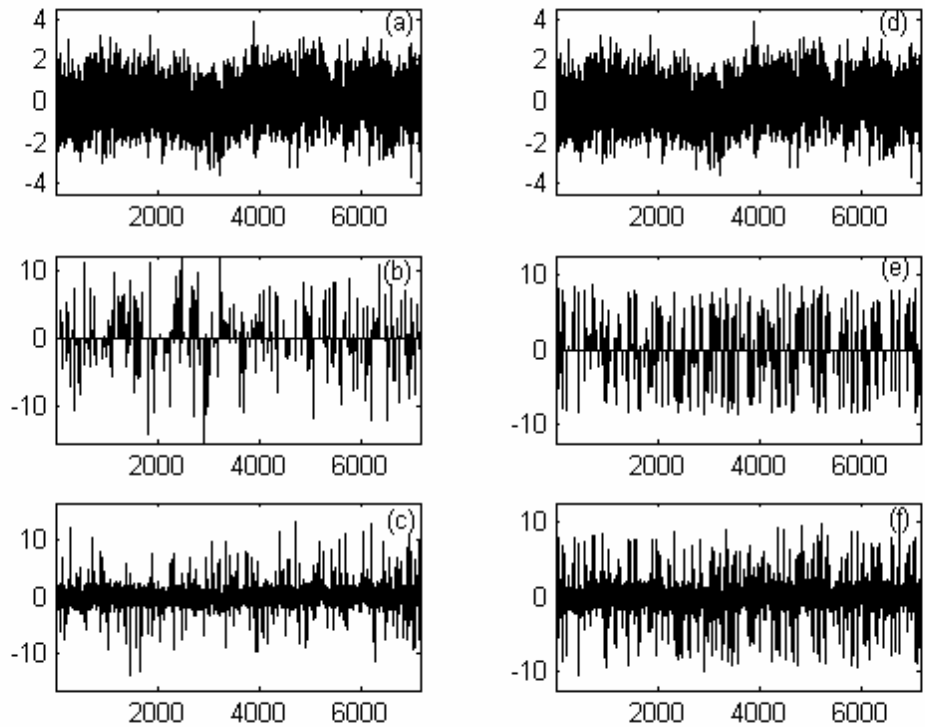

**Figure 4** Long range correlated noise (a, d) with ($\alpha = 0.8$, $\mu = 0$, $\sigma = 1$, $N = 7168$) superimposed with random spike generated from uncorrelated noise. Random spikes sampled from normally distributed uncorrelated noise (b) with ($\mu = 0$, $\sigma = 5$, $p = 0.05$) superimposed on the long-range correlated noise is shown in (c). Random spikes sampled from uniformly distributed uncorrelated noise (e) with ($\mu = 0$, $\sigma = 5$, $p = 0.05$) superimposed on the long-range correlated noise is shown in (f)

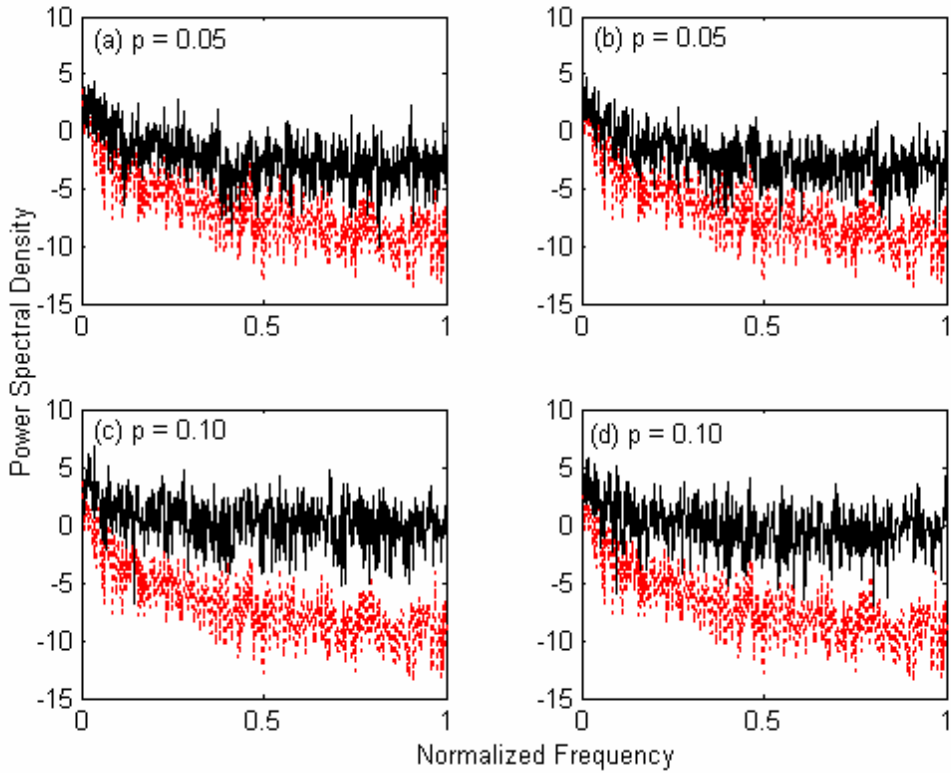

**Figure 5** Power-spectrum of the long-range correlated noise ($\alpha = 0.8$, $\mu = 0$, $\sigma = 1$, $N = 7168$) (dotted line) and that superimposed with random spike sampled from uncorrelated noise (solid line). (a) and (c) cases where the random spikes were sampled from normally distributed uncorrelated noise ($\mu = 0$, $\sigma = 5$) with acceptance probability ($p = 0.05$ and $0.10$). (b) and (d) represent cases where the random spikes were sampled from uniformly distributed uncorrelated noise ($\mu = 0$, $\sigma = 5$) with acceptance probability ($p = 0.05$ and $0.10$).

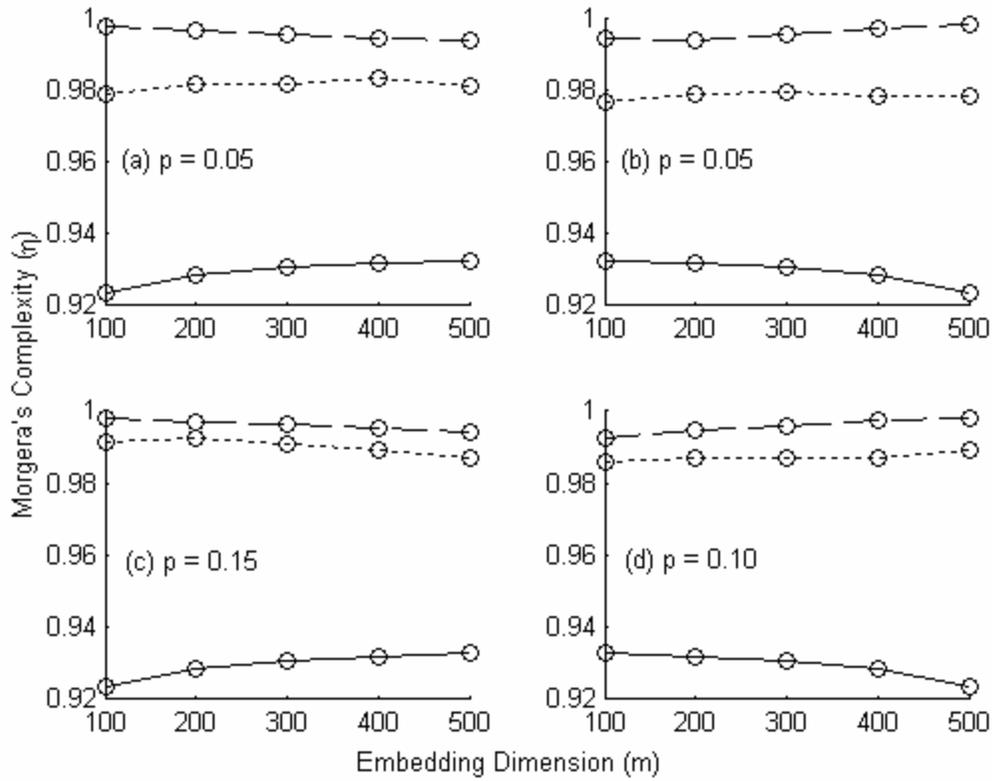

**Figure 6** Morgera's covariance complexity (*h*) of the long-range correlated noise ($\alpha = 0.8$, $\mu = 0$, $\sigma = 1$, $N = 7168$) (solid line), random spike (dashed lines) and long-range correlated noise superimposed with random spike (dotted lines) with varying embedding dimension (m = 100, 200, 300, 400 and 500). (a) and (c) represent cases where the random spikes were sampled from normally distributed uncorrelated noise ($\mu = 0$, $\sigma = 5$) with acceptance probability (p = 0.05 and 0.10). (b) and (d) represent cases where the random spikes were sampled from uniformly distributed uncorrelated noise ($\mu = 0$, $\sigma = 5$) with acceptance probability (p = 0.05 and 0.10).

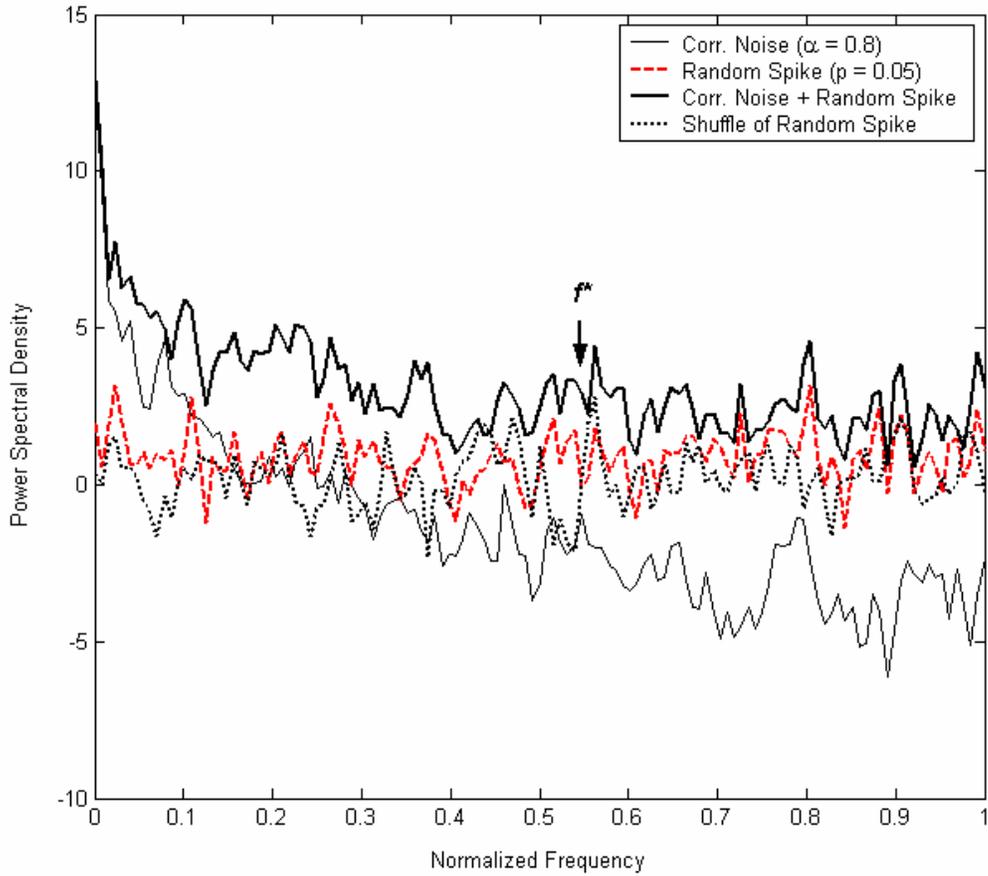

**Figure 7** Power-spectrum of the long-range correlated noise (α = 0.8, μ = 0, σ = 1, thin solid line), random spikes (dashed line) sampled from normally distributed uncorrelated noise (N = 7168, μ = 0, σ = 5, p = 0.05), long-range correlated noise superimposed with random spike (thick solid line) and its random shuffle (dotted line). The arrow indicated the point (*f\**) where the power long-range correlated noise superimposed with the random spikes for (*f* > *f\**) is dominated by random spikes.

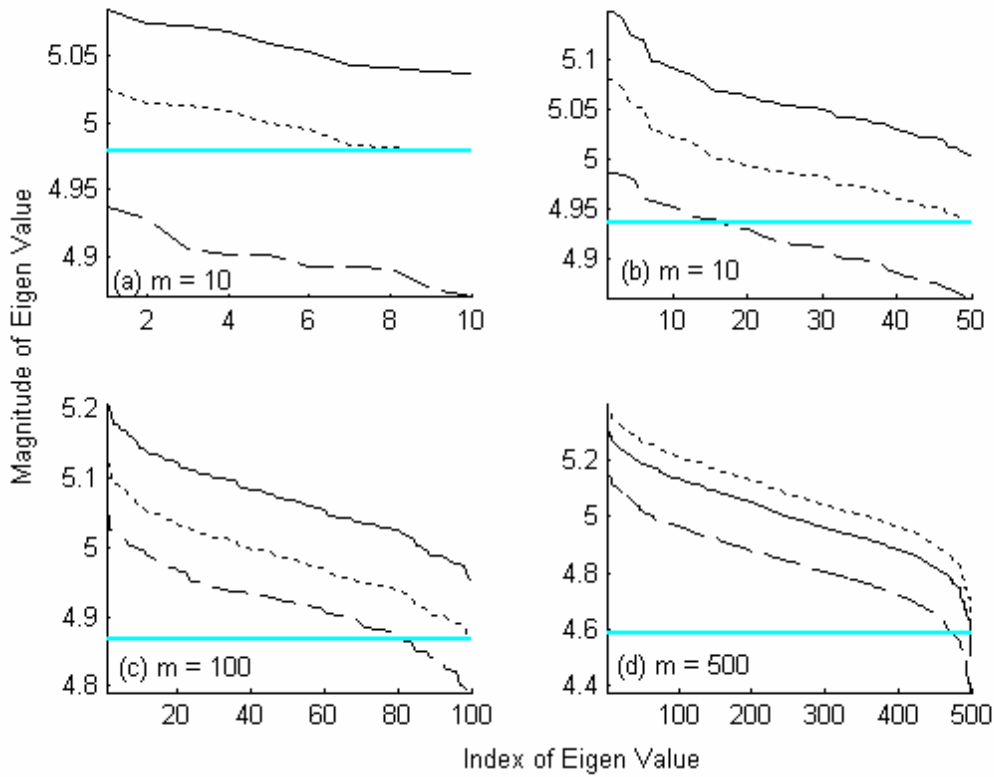

**Figure 8** Eigen-spectrum (log scale) of random spikes (dotted line) sampled from normally distributed uncorrelated noise ($\mu = 0$, $\sigma = 5$, $p = 0.10$), its random shuffled surrogate (solid line), scaled shuffled surrogate (dashed lines) and least-dominant eigen-value (solid flat line) corresponding to the superimposed data ($y = x + e$). Subplots (a, b, c, and d) correspond to embedding dimension (m = 10, 50, 100 and 500).

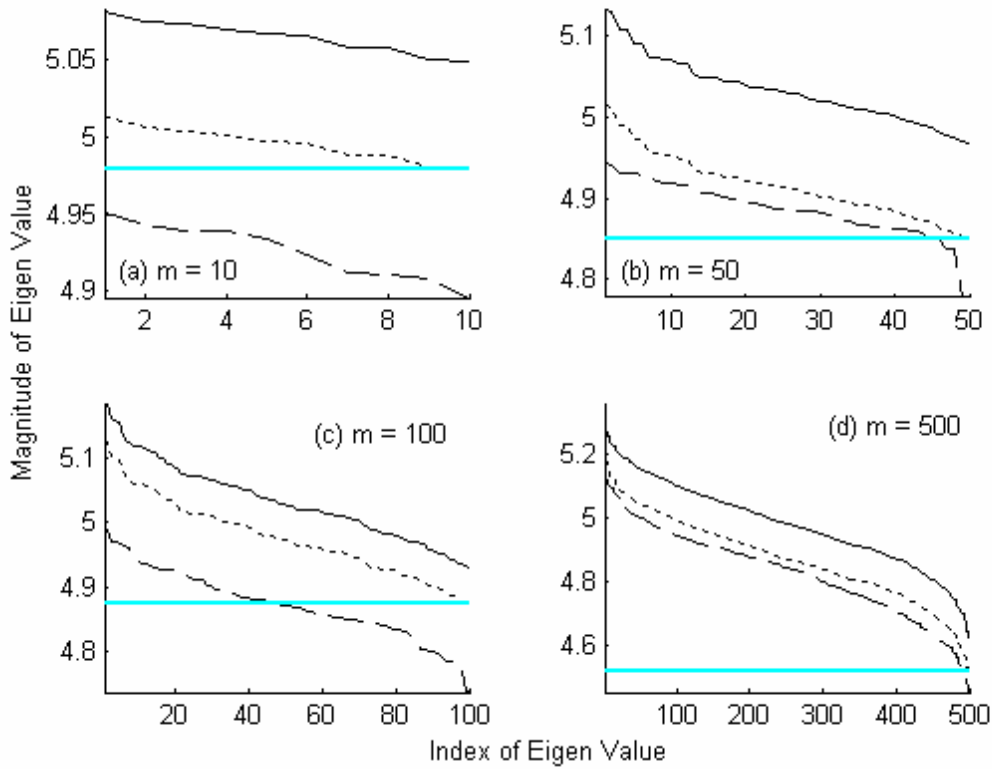

**Figure 9** Eigen-spectrum (log scale) of random spikes (dotted line) sampled from uniformly distributed uncorrelated noise ($\mu = 0$, $\sigma = 5$, $p = 0.10$), its random shuffled surrogate (solid line), scaled shuffled surrogate (dashed lines) and least-dominant eigen-value (solid flat line) corresponding to the superimposed data ($y = x + e$). Subplots (a, b, c, and d) correspond to embedding dimension (m = 10, 50, 100 and 500).

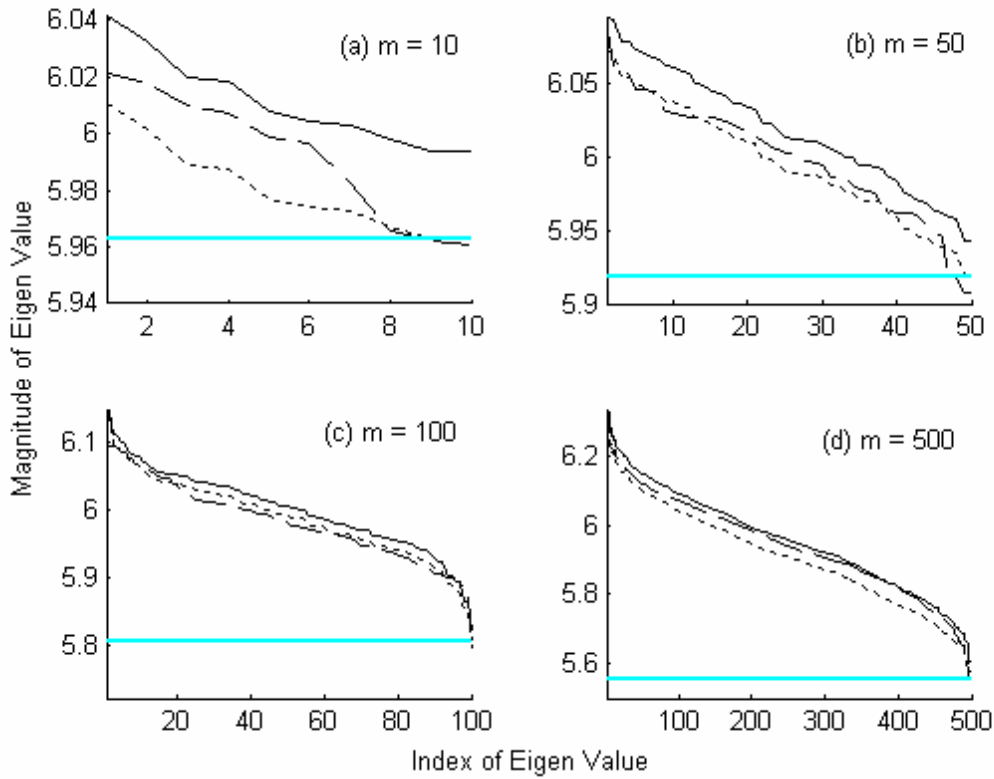

**Figure 10** Eigen-spectrum (log scale) of random spikes (dotted line) sampled from normally distributed uncorrelated noise (μ = 0, σ = 5, p = 0.90), its random shuffled surrogate (solid line), scaled shuffled surrogate (dashed lines) and least-dominant eigen-value (solid flat line) corresponding to the superimposed data (*y* = *x* + *e*). Subplots (a, b, c, and d) correspond to embedding dimension (m = 10, 50, 100 and 500).

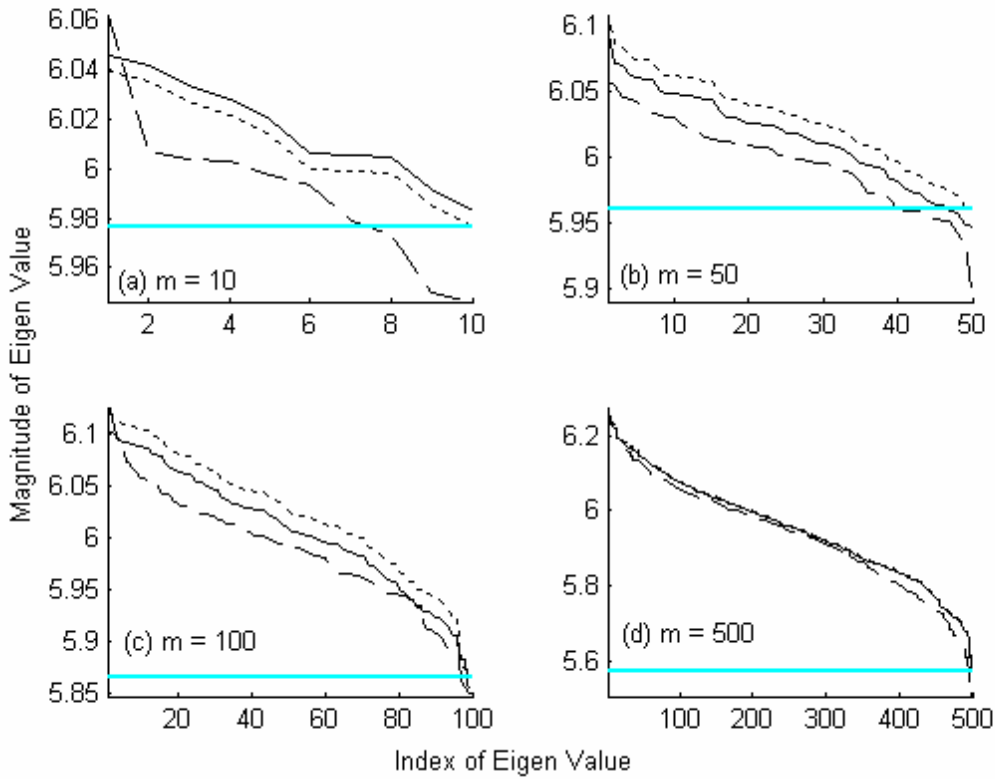

**Figure 11** Eigen-spectrum (log scale) of random spikes (dotted line) sampled from uniformly distributed uncorrelated noise ($\mu = 0$, $\sigma = 5$, $p = 0.90$), its random shuffled surrogate (solid line), scaled shuffled surrogate (dashed lines) and least-dominant eigen-value (solid flat line) corresponding to the superimposed data ($y = x + e$). Subplots (a, b, c, and d) correspond to embedding dimension (m = 10, 50, 100 and 500).

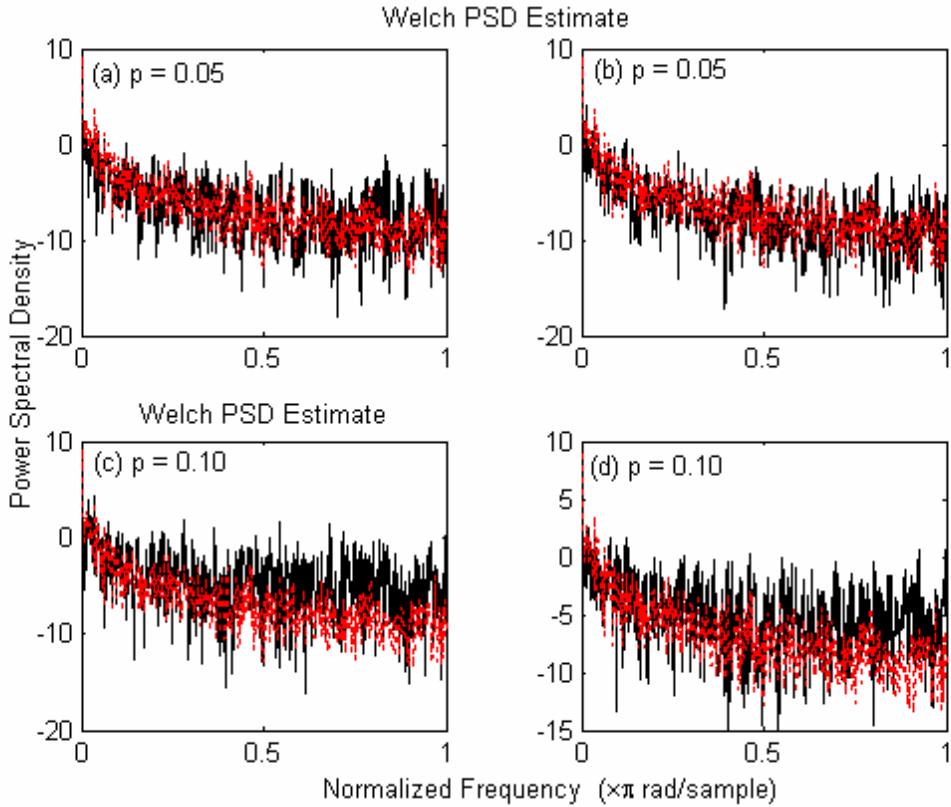

**Figure 12** Power-spectrum of the long-range correlated noise (α = 0.8, μ = 0, σ = 1,) superimposed with random spike generated uncorrelated noise after filtering with embedding dimension (m = 500). (a) and (c) represent random spikes sampled from normally distributed uncorrelated noise (μ = 0, σ = 5) with (p = 0.05 and 0.10) respectively. (b) and (d) represent random spikes sampled from uniformly distributed uncorrelated noise (μ = 0, σ = 5) with (p = 0.05 and 0.10) respectively. The power-spectrum of the long-range correlated noise (dotted lines) is shown inside each subplot for reference.

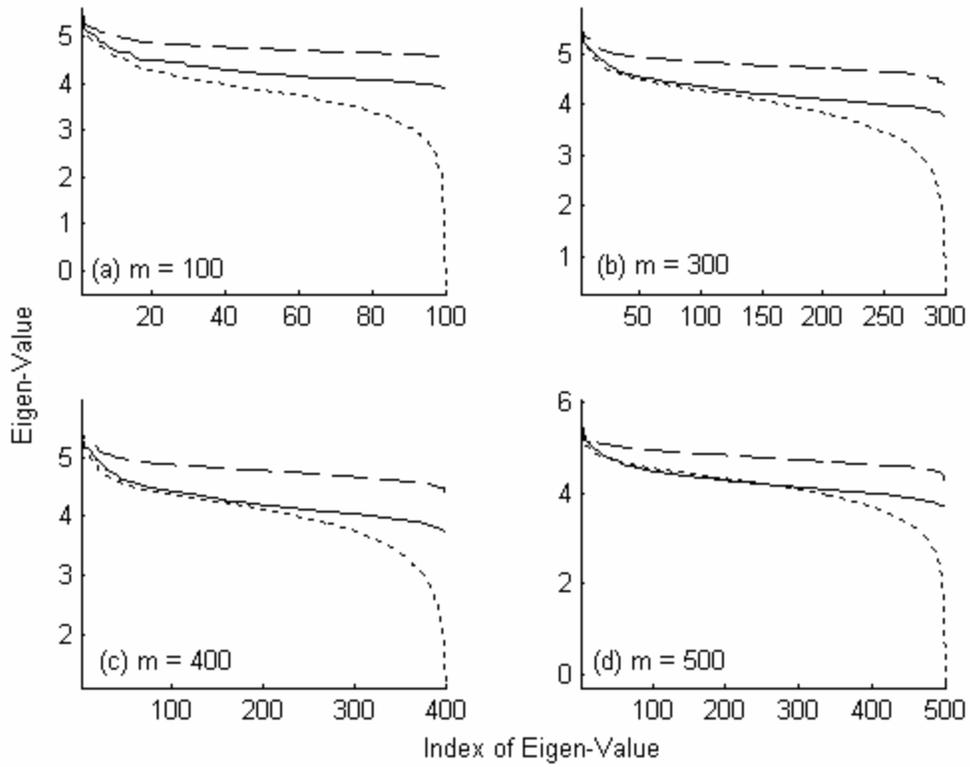

**Figure 13** Eigen-spectrum of the long-range correlated noise ($\alpha = 0.8$, $\mu = 0$, $\sigma = 1$, solid line), long-range correlated superimposed with random spikes sampled from normally distributed uncorrelated noise ($\mu = 0$, $\sigma = 5$, $p = 0.05$, dashed line) and the filtered data (dotted lines). The filtered data generated using ($m = 100, 300, 400$ and $500$ with $\tau = 1$) shown in (a, b, c, and d) respectively.

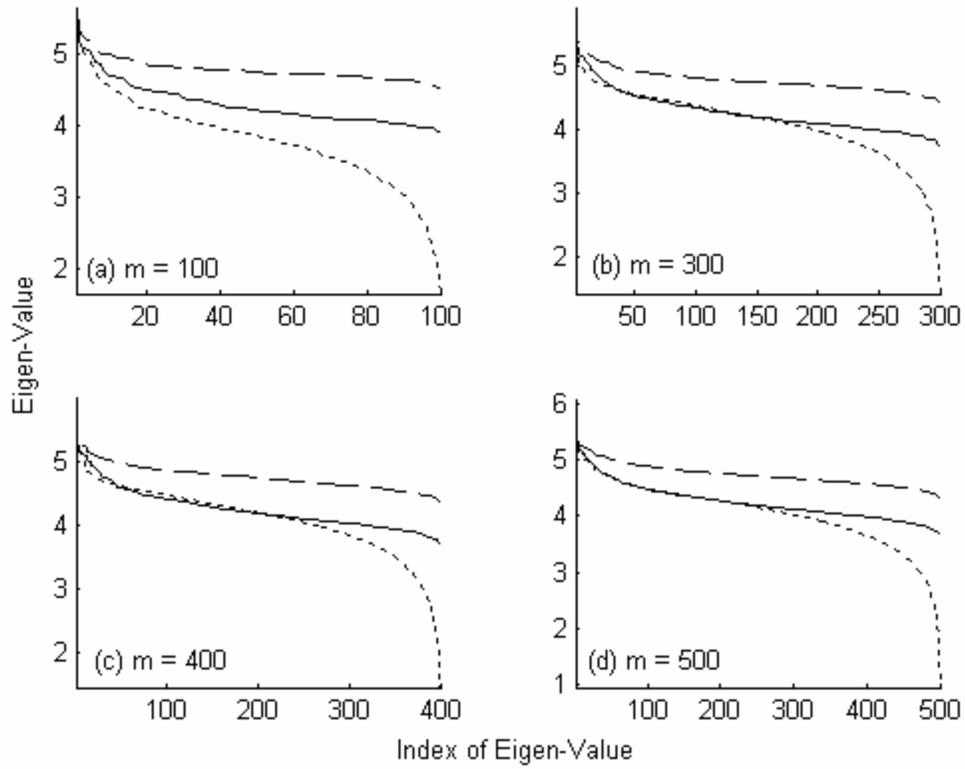

**Figure 14** Eigen-spectrum of the long-range correlated noise ($\alpha = 0.8$, $\mu = 0$, $\sigma = 1$, $N = 7168$, solid line), long-range correlated superimposed with random spikes sampled from uniformly distributed uncorrelated noise ($\mu = 0$, $\sigma = 5$, $p = 0.05$, dashed line) and the filtered data (dotted lines). The filtered data generated using (m = 100, 300, 400 and 500 with $\tau = 1$) shown in (a, b, c, and d) respectively.

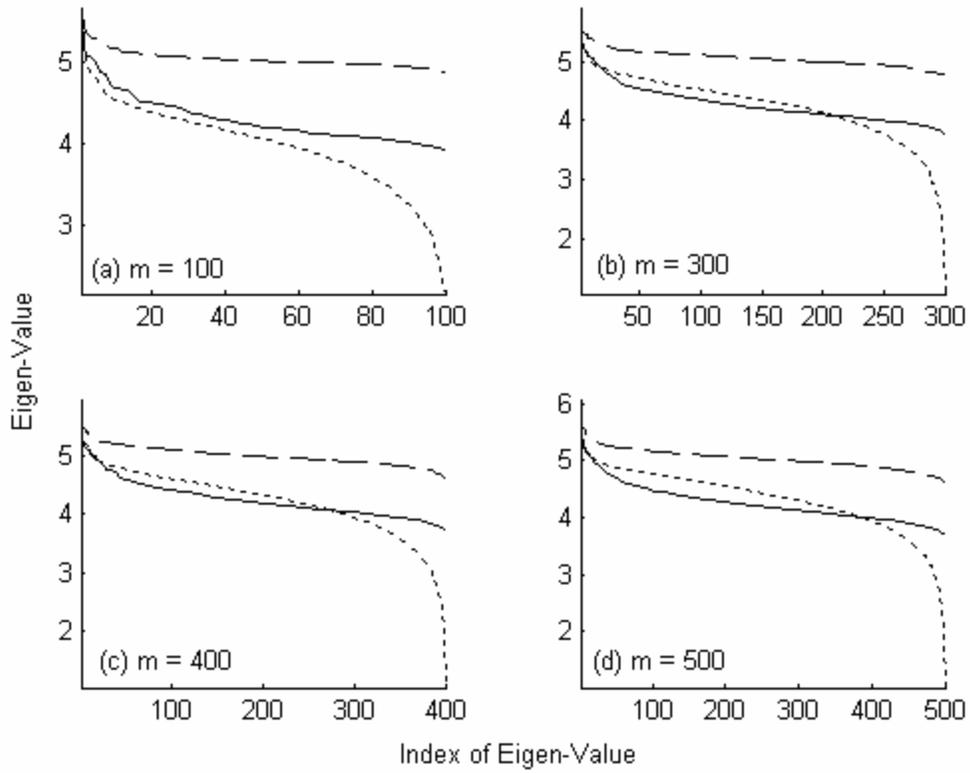

**Figure 15** Eigen-spectrum of the long-range correlated noise ($\alpha = 0.8$, $\mu = 0$, $\sigma = 1$, $N = 7168$, solid line), long-range correlated superimposed with random spikes sampled from normally distributed uncorrelated noise ($\mu = 0$, $\sigma = 5$, $p = 0.10$, dashed line) and the filtered data (dotted lines). The filtered data generated using ($m = 100, 300, 400$ and $500$ with $\tau = 1$) shown in (a, b, c, and d) respectively.

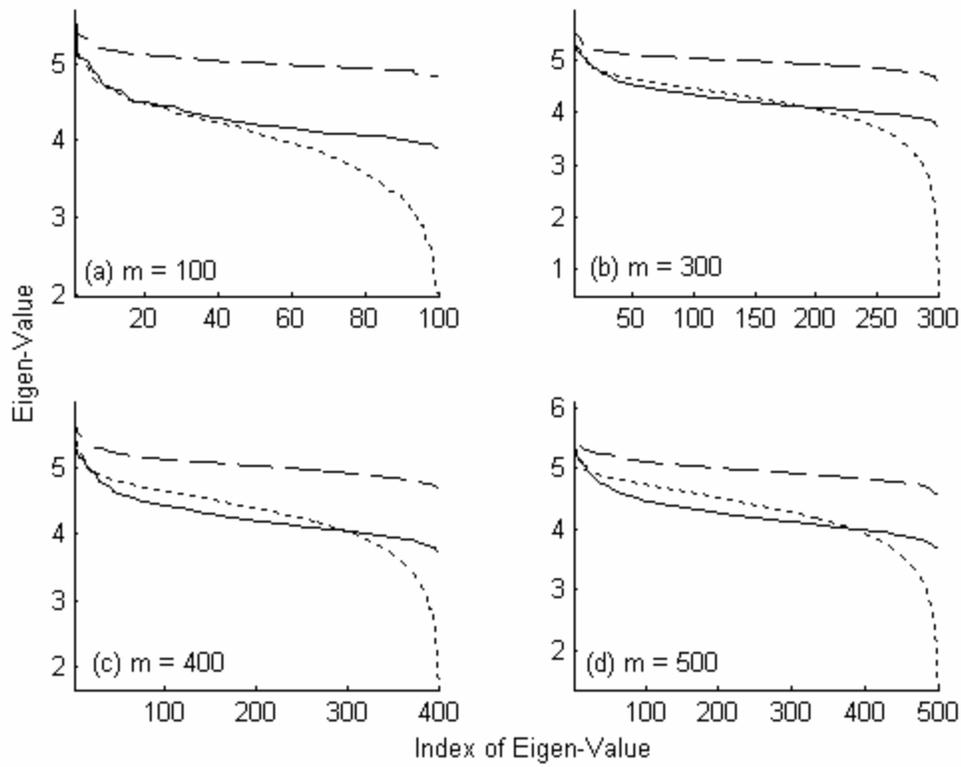

**Figure 16** Eigen-spectrum of the long-range correlated noise ($\alpha = 0.8$, $\mu = 0$, $\sigma = 1$, $N = 7168$, solid line), long-range correlated superimposed with random spikes sampled from uniformly distributed uncorrelated noise ($\mu = 0$, $\sigma = 5$, $p = 0.10$, dashed line) and the filtered data (dotted lines). The filtered data generated using ($m = 100, 300, 400$ and $500$ with $\tau = 1$) shown in (a, b, c, and d) respectively.

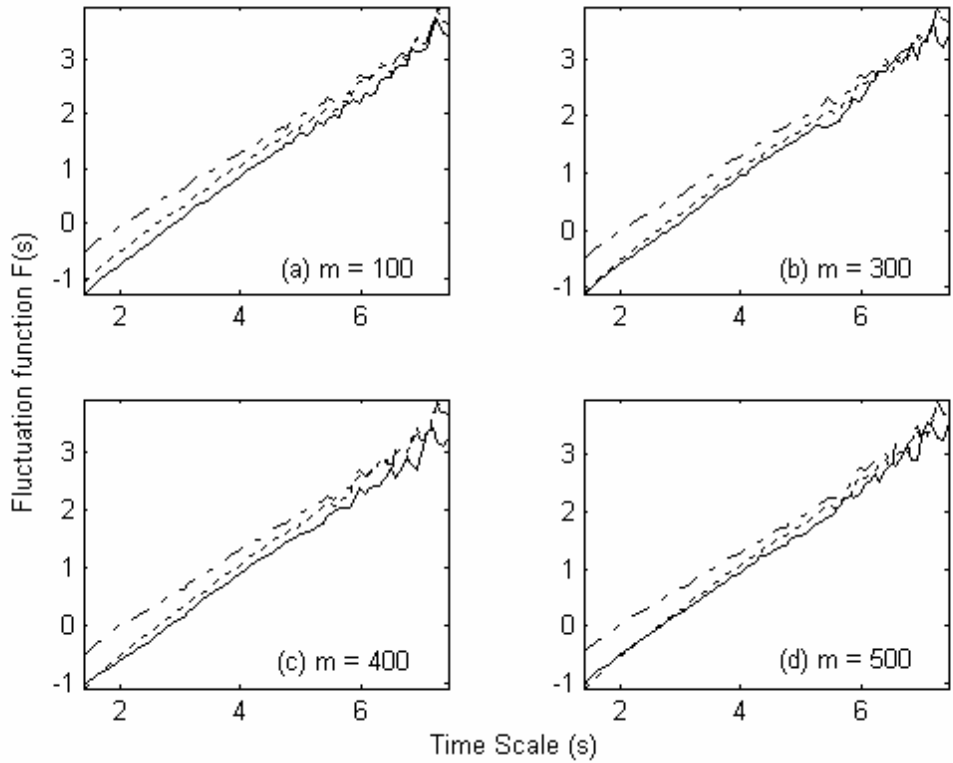

**Figure 17** Log-log fluctuation plots of the long-range correlated noise ($\alpha = 0.8$, $\mu = 0$, $\sigma = 5$, $N = 7168$, solid lines), long-range correlated superimposed with random spike sampled from normally distributed uncorrelated noise ($\mu = 0$, $\sigma = 5$, $p = 0.05$, dashed lines) and the filtered data (dotted lines) with varying embedding dimension ($m = 100, 300, 400$ and $500$ with $\tau = 1$) is shown in (a, b, c and d) respectively.

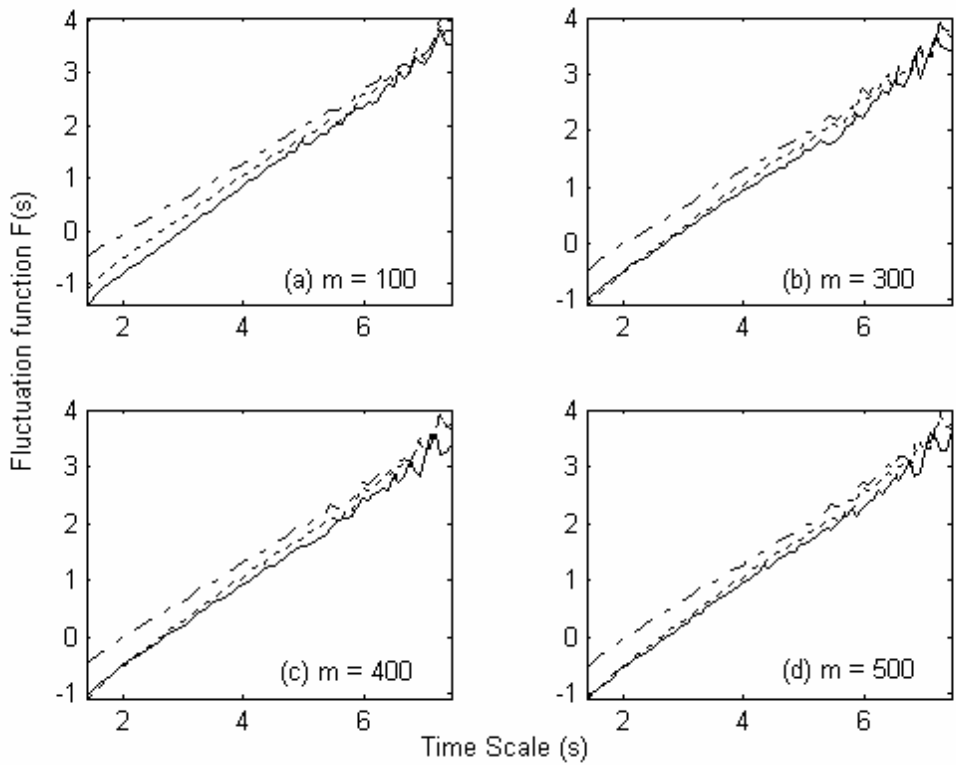

**Figure 18** Log-log fluctuation plots of the long-range correlated noise ($\alpha = 0.8$, $\mu = 0$, $\sigma = 5$, N = 7168, solid lines), long-range correlated superimposed with random spike sampled from uniformly distributed uncorrelated noise ($\mu = 0$, $\sigma = 5$, p = 0.05, dashed lines) and the filtered data (dotted lines) with varying embedding dimension (m = 100, 300, 400 and 500 with $\tau = 1$) is shown in (a, b, c and d) respectively.

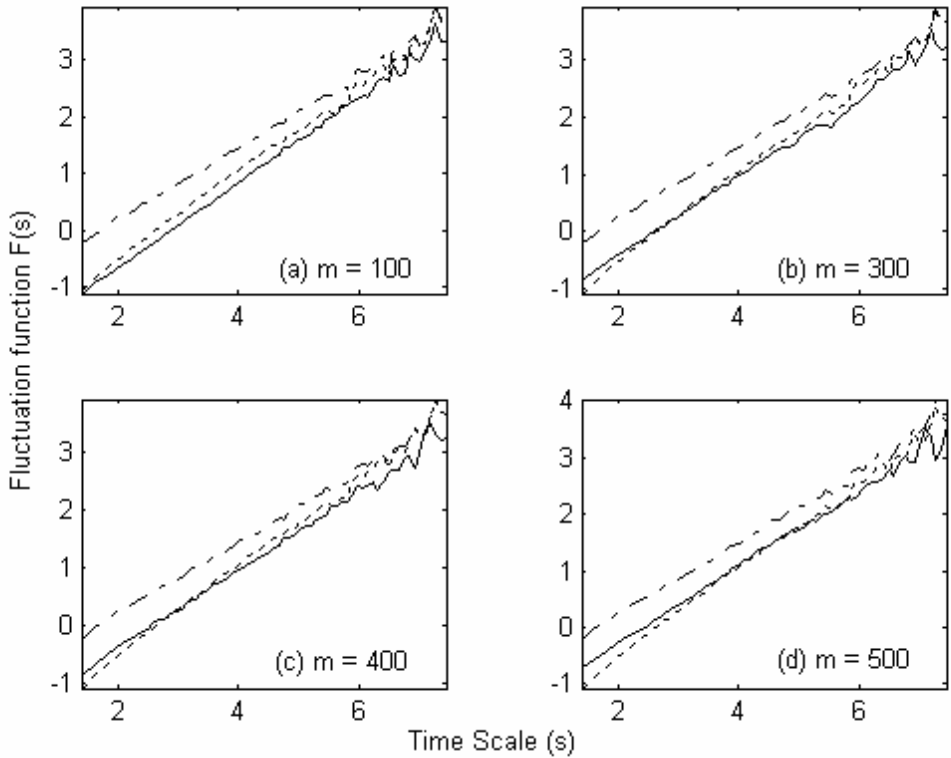

**Figure 19** Log-log fluctuation plots of the long-range correlated noise ($\alpha = 0.8$, $\mu = 0$, $\sigma = 5$, $N = 7168$, solid lines), long-range correlated superimposed with random spike sampled from normally distributed uncorrelated noise ($\mu = 0$, $\sigma = 5$, $p = 0.10$, dashed lines) and the filtered data (dotted lines) with varying embedding dimension ($m = 100, 300, 400$ and $500$ with $\tau = 1$) is shown in (a, b, c and d) respectively.

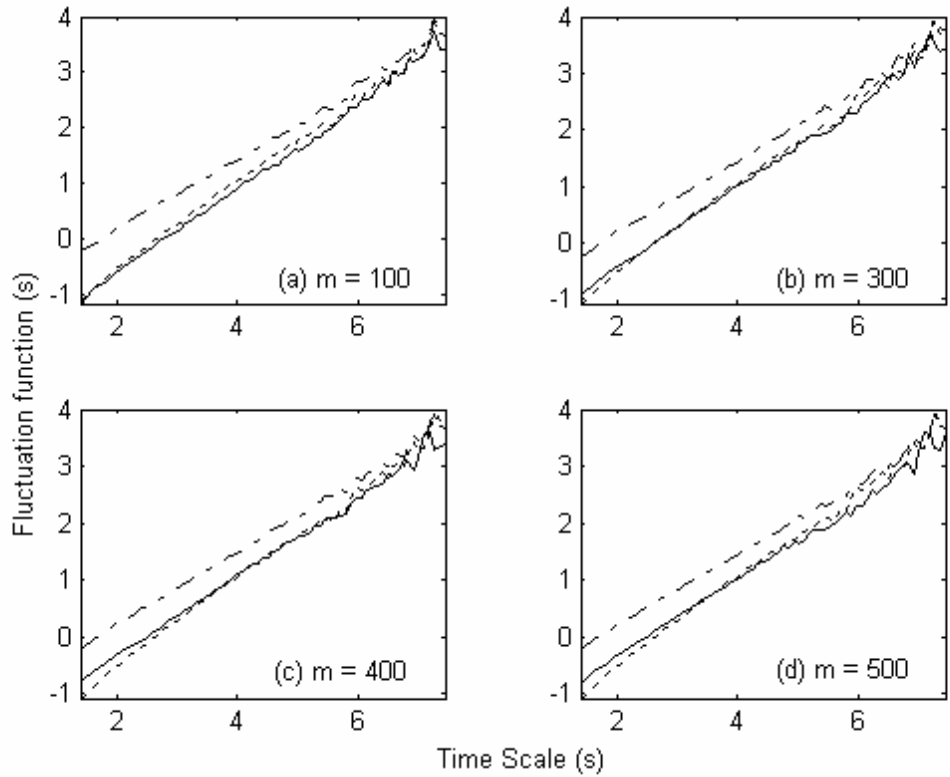

**Figure 20** Log-log fluctuation plots of the long-range correlated noise ($\alpha = 0.8$, $\mu = 0$, $\sigma = 5$, $N = 7168$, solid lines), long-range correlated superimposed with random spike sampled from uniformly distributed uncorrelated noise ($\mu = 0$, $\sigma = 5$, $p = 0.10$, dashed lines) and the filtered data (dotted lines) with varying embedding dimension ($m = 100, 300, 400$ and $500$ with $\tau = 1$) is shown in (a, b, c and d) respectively.